\documentclass[12pt]{article}
\usepackage{times}

\usepackage{amsmath}
\usepackage{amssymb}
\usepackage{float}
\usepackage{graphicx}

\usepackage{color} 

\usepackage{url}
\usepackage{setspace} 

\usepackage{pdfpages}
\usepackage{soul}
\usepackage{url}
\usepackage[hidelinks]{hyperref}
\usepackage[utf8]{inputenc}
\usepackage[small]{caption}
\usepackage{amsthm}
\usepackage{booktabs}
\usepackage{algorithm}
\usepackage{algorithmic}
\usepackage[comma,sort&compress]{natbib}
\usepackage{dblfloatfix}

\theoremstyle{definition}

\hypersetup{
	bookmarks=true,         
	unicode=false,          
	pdftoolbar=true,        
	pdfmenubar=true,        
	pdffitwindow=false,     
	pdfstartview={FitH},    
	pdftitle={My title},    
	pdfauthor={Author},     
	pdfsubject={Subject},   
	pdfcreator={Creator},   
	pdfproducer={Producer}, 
	pdfkeywords={keywords}, 
	pdfnewwindow=true,      
	colorlinks=true,       
	linkcolor=blue,          
	citecolor=blue,        
	filecolor=magenta,      
	urlcolor=cyan           
}

\usepackage[labelfont=bf,labelsep=period,justification=raggedright]{caption}

\makeatletter
\renewcommand{\@biblabel}[1]{\quad#1.}
\makeatother

\date{}

\title{Social Diversity Reduces the Complexity and Cost of Fostering Fairness}

\author{
Theodor Cimpeanu$^{1}$,
Alessandro Di Stefano$^1$,
Cedric Perret$^2$,\\
and The Anh Han$^{1,\star}$}

\begin{document}

\maketitle

	{\footnotesize
		\noindent
		$^{1}$ School Computing, Engineering and Digital Technologies, Teesside University\\
		$^{2}$  College of Life and Environmental Sciences, Exeter University\\ \\ 
		$^\star$ Corresponding author: The Anh Han (T.Han@tees.ac.uk)
	}

\newpage
\section*{Abstract}
Institutions and investors are constantly faced with the challenge of appropriately distributing endowments. No budget is limitless and optimising overall spending without sacrificing positive outcomes has been approached and resolved using several heuristics. To date, prior works have failed to consider how to encourage fairness in a population where social diversity is ubiquitous, and in which investors can only partially observe the population. Herein, by incorporating social diversity in the Ultimatum game through heterogeneous graphs, we investigate the effects of several interference mechanisms which assume incomplete information and flexible standards of fairness. We quantify the role of diversity and show how it reduces the need for information gathering, allowing us to relax a strict, costly interference process. Furthermore, we find that the influence of certain individuals, expressed by different network centrality measures, can be exploited to further reduce spending if minimal fairness requirements are lowered. Our results indicate that diversity changes and opens up novel mechanisms available to institutions wishing to promote fairness. Overall, our analysis provides novel insights to guide institutional policies in socially diverse complex systems. 


 \vspace{0.2in}
 
 \noindent \textbf{Keywords:} fairness, cost efficiency, decision making, ultimatum game, social diversity, evolutionary game theory.

 \newpage


\section{Introduction}

\textit{Fairness} has a deep impact on decision-making and individuals often prefer fair outcomes over payoff-maximising ones \citep{nowak2000fairness,rand2013evolution}. For example, fairness concerns emerge and play a crucial role in group interactions, when agents must decide upon outcomes possibly favouring different parts
unequally \citep{teixeira2021evolution}. These concerns arise in many domains -- hybrid collectives of humans and machines \citep{paiva2018engineering}, wildlife management \citep{levin2000multiple}, conflict resolution \citep{pritchett2017negotiated} or enforcing global climate change actions \citep{gois2019reward, Ostrom2010, Pacheco2014}, just to name a few. In this context, several mechanisms have been identified to explain why fairness is widespread in human decision-making, but it is typically assumed to emerge from the actions of individuals within the system. 
Whether due to risk or safety concerns, there exist several domains, such as the ones listed above, that call for a new set of heuristics aimed at promoting fair outcomes in populations of self-regarding individuals.

Humans have developed considerable machinery used at scale to create policies and to distribute incentives, yet we are always searching for ways in which to improve upon these organisations, often referred to as  institutions \citep{ostrom1990governing, North1990}.  In these scenarios, external decision-makers must find a trade-off between the cost of the investment and its effectiveness in ensuring high levels of fair behaviour.
Several works have provided insights on how best to promote cooperation while also considering the costs of such \textit{interference} \citep{han2018cost,chen2014optimal, DuongHanPROCsA2021, cimpeanu2021cost,wang2019exploring}. Fairness, modelled using the Ultimatum Game \citep{nowak2000fairness}, has seen relatively little attention in the literature, and previous works have so far only considered an ideal world in which interactions are perfectly homogeneous \citep{cimpeanu2021cost}. Nevertheless, real-world networks of individuals, such as social networks and networks of collaboration,  are inherently \textit{heterogeneous}  \citep{barabasi1999emergence}. Moreover, in the context of  Evolutionary Game Theory (EGT), scale-free networks imply more than the underlying interaction structure. Heterogeneous graphs can portray social diversity \citep{santos2008social}, and the inherent inequality that exists between individuals. Individuals vary in influence and accumulated wealth, and this lends diversity particular importance in the quest towards fairness for two main reasons. Firstly, it has been shown to play a key role in the evolution of cooperative behaviours \citep{santos2006pnas,poncela2007robustness} and the emergence of fairness \citep{sinatra2009ultimatum}. It may enhance the resilience of cooperation, inducing cooperative agents to create assortative clusters, where they reciprocate cooperation \citep{santos2006pnas,di2020novel,di2015quantifying}. Secondly, decision-makers can then base their investment strategies not only on the state of the system at a global scale, but also on individual characteristics found only in certain parts of the network. This could potentially allow for novel interference strategies, which would maintain high standards of fairness at a reduced cost. For instance, an institution might decide to focus its efforts on building up deprived neighbourhoods or selectively invest only in very influential samaritans.



 Herein  we systematically investigate several interference mechanisms  for promoting the evolution of fairness in heterogeneous network settings assuming incomplete information and flexible standards of fairness.
We resort to the Ultimatum Game \citep{nowak2000fairness} as a suitable mathematical approach to model fair decision making (see further details in Section \ref{sec:ug}).  In the Ultimatum Game, one of the players can decide how to split a sum of money. Offers close to an even split are considered fair. An uneven split, in which the proposer gets to keep most of the money, is considered unfair. Because the proposer has asymmetric power in the interaction, the only way in which they can be "punished" is if the responder declines an unfair offer, thus causing neither individual to receive anything from the original sum. Fair individuals would always propose an even split, and always decline unfair offers to prevent selfish players from receiving part of the endowment. In contrast, very unfair individuals would propose to keep most if not all of the endowment while accepting anything they are given. 
We determine how these heterogeneous network characteristics can be exploited to reduce costs while maintaining high standards of fairness, providing insights on how the presence of \textit{social diversity} alters the complexity of engineering fairness and how it can be done efficiently.

Optimising fairness becomes especially challenging when studying evolving populations that incorporate diverse stochastic effects and uncertainty factors, such as a non-deterministic behaviour update. Undesired behaviours can reoccur over time and in order to hinder them, external decision-makers must repeatedly interfere in the system. Our focus is therefore on \textit{cost-efficient interference} problems, namely exploring how efficiently to interfere in a spatially heterogeneous population to achieve high levels of fairness while minimising the cost of interference. 
Given its relevance, this cost-efficient interference problem has naturally attracted significant attention in both evolution of collective behaviours  and computational modelling research. It can be formulated as a bi-objective optimisation problem, where the target is dual -- finding a cost-efficient interference scheme leading to the desired goal, while minimising the total cost of interference \citep{han2018cost,wang2019exploring,chen2015first,cimpeanu2021cost}. 

Our results show that social diversity reduces the complexity and cost of promoting the evolution of fairness in several respects. Positive outcomes can be achieved without extensive information gathering, by reducing the strictness of eligibility for a player to receive an incentive. Moreover, lowering the required standards of fairness allows decision-makers to capitalise on the influence of hubs to reduce costs. Finally, we find that heterogeneous networks have a tendency towards polarisation, and suggest that this can be exploited to induce pro-sociality more efficiently.

\section{Models and Methods}
\label{sec:methods}
\subsection{Ultimatum Game (UG)}
\label{sec:ug}
In our work, the interaction between agents is modelled using the one-shot Ultimatum Game (UG) \citep{nowak2000fairness,page2000spatial}, which serves as a useful framework for addressing and exploring the evolution of fairness. In this game, two players are asked to divide a sum and possibly win a certain amount of money. 
One player (the \textit{proposer}) suggests how to split the sum. The other player (the \textit{responder}) decides whether to accept or reject the offer. Only if the responder accepts, will the sum be shared as proposed. Otherwise, neither player receives any part of the initial sum. 
Coherently with \citep{nowak2000fairness,page2000spatial}, we assume that a player is equally likely to act in either role. A player's strategy is defined by a pair of parameters, $p$ and $q$, where $p$ is the proposer's offer, while $q$ is the acceptance threshold. That is, when acting as proposer, the player offers $p$, whereas in a receiver's role, the player rejects any offer smaller than $q$. 

In this paper, we focus on how the presence of multiple roles in the interactions affects decision making in the investment process. As such, we consider a baseline UG model where proposers have two possible strategic offers, a low (L, with  $p=l$) and a high one (fair) (H, with  $p = h$), where $l < h$, with $l, h \in [0,1]$. On the other hand, receivers have two options, a low threshold (L, with $q = l$) and a high threshold (H, with $q = h$). Thus, overall, there are four possible strategies HH, HL, LH and LL (e.g., HL denotes a strategy that offers high and accepting any offers).
\begin{figure}[ht]
    \centering
    \includegraphics[width=1.\linewidth]{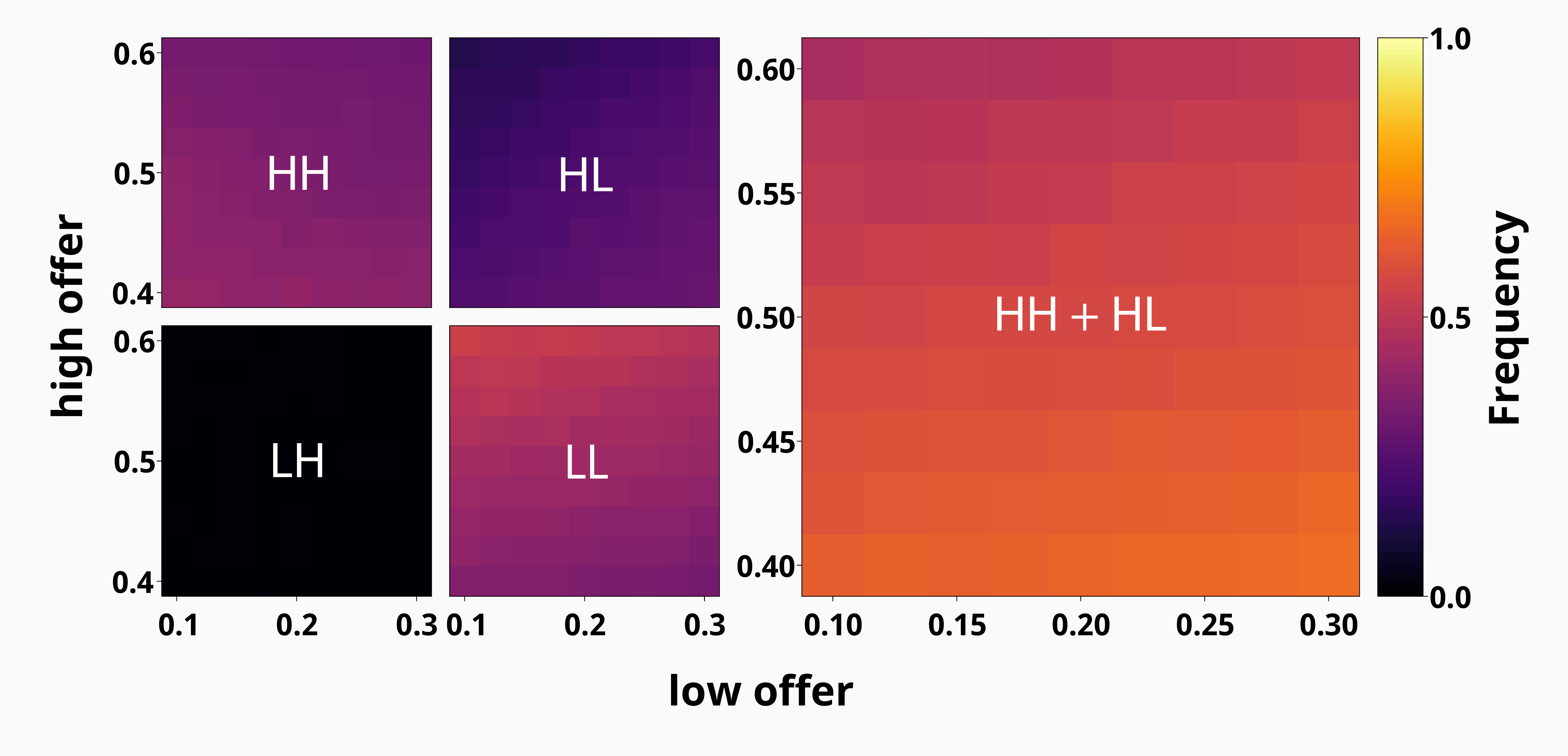}
    \caption[Baseline frequencies]{\textbf{Baseline frequencies}. The figure shows the baseline frequencies for each strategy for scale-free (BA) networks, with a separate panel for overall fairness (fair offers). Colour corresponds to the frequency, ranging from ‘black’ (lowest) to ‘yellow’ (highest) (see colour bar). The figure shows the frequency of offers in a LH (Low-High offers) plane, which is divided into four quadrants representing the different strategies: HH (upper-left), HL, (upper-right), LH (lower-left), and LL (lower-right).}
    \label{fig:ug-baseline-ba}
\end{figure}

In our model of the UG, \textit{fairness} is measured by calculating what percentage of the population is representative for either the HH or HL strategies (i.e., fair proposers). This allows us to have a clear comparison with previous works---in terms of fairness achieved at the level of population---that have investigated the evolution of fairness in the UG, such as in \citep{page2000spatial,nowak2000fairness,rand2013evolution}. 



Particularly, we set $h = 0.6$ and $l = 0.1$, as this represents the environment with (roughly) the lowest frequency of fair proposals (see Figure \ref{fig:ug-baseline-ba}; for DMS networks,  see Figure S1 in Supporting Information (\textbf{SI})). Furthermore, in line with evidence from behavioural experiments \citep{guth1982experimental,rand2013evolution}, which suggests that people (almost) never offer more than half of the sum. We also explore the typical case of $h = 0.5$ and validate our findings (see Figures S2 and S3 in  SI). 

\newpage
The payoff matrix  for the four strategies HH, HL, LH, and LL reads (for row player):
\begin{center}	
\begin{table}[h]
\centering
\begin{tabular}{c|c c c c c}
& HH & HL & LH & LL \\
\toprule
HH & $\frac{1}{2}$ & $\frac{1}{2}$ & $\frac{1 - h}{2}$ & $\frac{1 - h}{2}$ \\
HL &  $\frac{1}{2}$ & $\frac{1}{2}$ &  $\frac{1 - h + l}{2}$ & $\frac{1 - h + l}{2}$ \\
LH & $\frac{h}{2}$ &  $\frac{1 + h - l}{2}$ & $0$ & $\frac{1 - l}{2}$ \\
LL & $\frac{h}{2}$ &  $\frac{1 + h - l}{2}$ & $\frac{l}{2}$ & $\frac{1}{2}$ 
\end{tabular}
\end{table}
\end{center}
For example, an HH player encountering an HL player results in the payoff $\frac{1}{2}$ for either player, as both of them propose and accept a fair split (i.e.,one interaction results in the payoff $1 - h$ for the proposer, and $h$ for the receiver, and vice-versa for when the roles are reversed). 

\subsection{Population structure and evolutionary  dynamics}

Initially each agent is designated as one of the four strategies (i.e., HH, HL, LH, HH), with equal probability. 
At each time step, each agent plays the UG with its immediate neighbours. The score for each agent is the sum of the payoffs in these encounters. At the end of each step, an agent $A$ with fitness $f_A$ chooses to copy the strategy of a randomly selected neighbour agent $B$ with score $f_B$, with a probability given by the Fermi function (i.e., \textit{stochastic update}) \citep{szabo2007evolutionary}: $$W(S_{B}\rightarrow S_{A})=(1+e^{(f_A- f_B)/K})^{-1},$$ where $K$ denotes the amplitude of noise in the imitation process \citep{szabo2007evolutionary}. In line with previous works and lab experiments \citep{szabo2007evolutionary,rand2013evolution}, we set $K = 0.1$ in our simulations. 

We simulate this evolutionary process until a stationary state or a cyclic pattern is reached. 
For the sake of a clear and fair comparison, all simulations are run for 500000 generations. 
Moreover, for each simulation, the results are averaged over the final 25000 generations, in order to account for the fluctuations characteristic of these stable states. 
Furthermore, to improve accuracy, for each set of parameter values, the final results are obtained from averaging 20 independent realisations. When shown in figures, the error bars represent the standard error of the mean between replicates.


\subsection{Network Topologies}

The Barab{\'a}si and Albert (\textbf{BA}) model \citep{barabasi1999emergence} is one of the most famous models used in the study of highly heterogeneous, complex networks. The main features of the BA model are that it follows a \textit{preferential attachment} rule, has a small clustering coefficient, and a typical \textit{power-law degree distribution}. In order to explain preferential attachment, let us describe the construction of a BA network. Starting from a small set of $m_0$ interconnected nodes, each new node selects and creates a link with $m$ older nodes according to a probability proportional to their degree. The procedure stops when the required network size of $N$ is reached. This will produce a network characterised by a power-law distribution, $p_k \sim k^{-\gamma}$, where the exponent $\gamma$ is its $\textit{degree exponent}$ \citep{Barabasi2016}. There is a high degree correlation between nodes, and the degree distribution is typically skewed with a long tail. There are few hubs in the network that attract an increasing number of new nodes which attach as the network grows (in a typical \textit{``rich-get-richer''} scenario). The power-law distribution exhibited by BA networks resembles the heterogeneity present in many real-world networks, however, they are also defined by low clustering coefficients, which means they cannot always be used to approximate realistic settings \citep{Su2016}.

To build heterogeneous networks with a large clustering coefficient, \citet{dorogovtsev2000structure} have proposed the eponymous Dorogovtsev-Mendes-Samukhin (\textbf{DMS}) model. This model follows a similar method of construction as the BA model. The crucial difference is that each new node connects with the two extremities of $m$ ($m \geq 2$) randomly chosen edges, instead, therefore forming characteristic triangular motifs whenever a new node is added to the network. Since the number of edges arriving to any node reflects its degree, the probability of attaching the new node to an old node is proportional to its degree and preferential attachment is recovered. The degree distribution is therefore the same as the one of a BA model, and the degree-degree correlations are also equal \citep{dall2006nonequilibrium}. However, the clustering coefficient is large, and more accurately mimics many realistic social networks \citep{Su2016,barrat2005rate}. The average connectivity for both types of scale-free networks is $z = 2m$. For all of our experiments, we seed 10 different networks (of each type) of size $N = 2000$, with an average connectivity of $z = 4$.

\subsection{Cost-Efficient Interference in Networks}
We aim to study how one can efficiently interfere in a spatially heterogeneous population to achieve high levels of fairness while minimising the cost of interference. As mentioned above, the level of fairness is measured by the fraction of fair offers in the population \citep{rand2013evolution}.
An investment decision  consists of a cost  $\theta > 0$ to the external decision-making agent/investor, and this value $\theta$ is added as surplus to the payoff of each suitable candidate. In order to determine cost-efficiency, we vary $\theta$ for each proposed interference strategy, measuring the total accumulated costs to the investor. Thus, the most efficient interference schemes will be the ones with the lowest relative total cost.

We examine  different approaches to interference, where  fairness is advocated for either role or both, leading to different desirable behaviours to be targeted:
\begin{itemize} 
    \item[\textbf{(i)}] ensure all proposals are fair, thus investing in  HH and HL (\textbf{Target: HH, HL});
    \item[\textbf{(ii)}] ensure only fair offers are accepted, thus investing in HH and LH (\textbf{Target: HH, LH}); 
    \item[\textbf{(iii)}] ensure both (i) and (ii), i.e., investing  in HH only (\textbf{Target: HH}). 
\end{itemize} 
Moreover, in line with previous works on network interference \citep{chen2015first,han2018ijcai,han2018cost,cimpeanu2019exogenous},  we compare  global  interference strategies where investments are triggered based on network-wide information, local neighbourhood information, and, lastly, node centrality information. 

In the \textit{population-based} (\textbf{POP}) approach,  a decision to invest in desirable behaviours is based on the current composition of the population. We denote $x_f$ the  fraction  of individuals  in the population with a desirable behaviour, given a targeting approach, i.e., (i), (ii) or (iii) as defined above. Namely, an investment is made if $x_f$ is less or at most equal to a threshold $p_f$ (i.e., $x_f \leq p_f$), for $0 \leq p_f \leq 1$. They do not invest otherwise (i.e., $x_f > p_f$).  The value $p_f$ describes how rare the desirable behaviours should be to trigger external support.  

In the \textit{neighbourhood-based} (\textbf{NEB}) approach, a decision to invest is based on the fraction $x_f$ of neighbours of a focal individual with the desirable behaviours, calculated at the local level. Investment happens if $x_f$ is less or at most equal to  a threshold $n_f$ (i.e., $x_f \leq n_f$), for $0 \leq n_f \leq 1$; otherwise, no investment is made.

As the presence of structural heterogeneity  in scale-free networks introduces a level of inequality between nodes in terms of influence, we also examine a \textit{node-influence-based} (NI) approach. 
 Here, we build upon the literature by introducing two measures for defining a node's influence, \textit{degree centrality} (NI-deg) and \textit{eigenvector centrality} (NI-eig). 

Firstly, degree centrality is the oldest measure of influence used in network science \citep{boldi2014axioms}. It denotes the number of neighbours of the node $i$ (i.e., its number of incoming edges). By definition, degree centrality is normalised using the total number of nodes, or the maximal possible degree, $n-1$, to obtain a number between $0$ and $1$. \textit{Degree centrality}, denoted by \textbf{NI-deg} or $x_{i}^{deg}$, is defined as follows:
\begin{equation}
    x_{i}^{deg}=deg_{i}=\frac{k_i}{n-1},
\end{equation}
where $k_{i}$ is the degree of the node $i$ and $n-1$ is the total number of nodes. The degree $k_i$ of a node $i$ is given by: $k_i = \sum_{j=1}^{n} A_{ij}$.

Despite its simple definition, degree centrality is often a highly effective measure of the influence or importance of a node, since people with more connections tend to be more influential in a social network \citep{bloch2019centrality}. 
However, degree centrality lacks potentially important aspects of the network's architecture or the node's position in the network. It can be considered a `local' centrality measure, measuring only the quantity and not the quality of connections \citep{di2015quantifying,di2020novel}.

Secondly, eigenvector centrality represents a related measure of prestige, since the importance of a node $i$ depends on the prestige of its neighbours \citep{bloch2019centrality, perra2008spectral}. In other words, this centrality measure acknowledges that not all connections are equal, but connections to nodes who are themselves influential will make a node more influential \citep{perra2008spectral}. Eigenvector centrality is computed by assuming that the centrality of node $i$ is proportional to the sum of the centralities of node $i$'s neighbours. Thus, \textit{eigenvector centrality}, denoted by \textbf{NI-eig} or $x_{i}^{eig}$, is defined as follows:
\begin{equation}
    x_{i}^{eig}=eig_{i}=\frac{1}{\lambda}\sum_{j=1}^{n} A_{ij} x_{j},
\end{equation}
where $\lambda$ is a positive constant or proportionality factor.

Defining the vector of centralities $\mathbf{x} = (x_1, x_2, \dots, x_n)$, we can rewrite the previous definition in matrix form as $\lambda \cdot \mathbf{x} = \mathbf{A} \cdot \mathbf{x}$. Hence $\mathbf{x}$ is an eigenvector of the adjacency matrix with eigenvalue $\lambda$. Since the centralities must be non-negative, it can be shown (using the Perron–Frobenius theorem) that $\lambda$ must be the largest eigenvalue of the adjacency matrix and $\mathbf{x}$ the corresponding eigenvector. 
Thus, even if a large number of connections increases the centrality measure, a node with a smaller number of high-quality nodes may still outrank one with a larger number of low-influential nodes. 

We denote by $x_i$  the node's centrality measure (e.g., for NI-deg: $x_i = deg_{i}=x_{i}^{deg}$). The nodes are sorted in ascending order based on their influence $x_i$, and the threshold $i_f$ denotes the fraction of nodes that will be selected for interference, if their behaviour satisfies the given targeting approach. For instance, given a network of size 1000 and a threshold $i_f = 0.001$ would mean selecting only the most influential node in the network for investment. 

Irrespective of the interference scheme or the targeted behaviour, the threshold signifies an increase in the number of nodes that satisfy the requirements for investment. In other words, a threshold of 1 means investing in all nodes which follow the desired strategy. Conversely, a lower threshold implies a more careful approach to investment, whereby the exogenous agent is stricter in their selection of suitable candidates. Moreover, the target selection also affects the number of candidate nodes. Stricter schemes, such as targeting individuals who are fair when proposing, and also when responding (HH), narrow the search for nodes which satisfy the requirements.

\section{Results}
\label{sec:results}

Several factors must be taken into consideration before deciding to invest in a population of individuals in order to adequately promote fairness. Among these, we take into account and resolve questions related to the importance of different roles of players, the size of the endowments, the threshold at which to resume investment, and the amount and quality of information that is available to the external decision-maker. We also consider that there exists a \textit{minimal level of fairness} which the external decision-maker aims to enforce \citep{han2018cost, han2018ijcai, cimpeanu2021cost}, and we study the most efficient strategies according to these varying standards of fairness. 

We structure the subsections that follow according to key insights derived from the results, and refer to previous results on structured populations \citep{cimpeanu2021cost}. We will highlight the key differences that arise in the presence of diversity (in the form of spatial heterogeneity), and mention similarities where appropriate.  

\begin{center}
\begin{table*}
\centering
 \caption[Most cost-efficient interference schemes to reach a minimum fairness of proposals in BA networks]{\textbf{Most cost-efficient interference schemes to reach a minimum fairness of proposals in BA networks}. For each minimal standard of fairness, we highlight (in bold) the least costly options across schemes.} 
 \label{table:ug-ba-full}
\small
\begin{tabular}{c c c c c c}\toprule
Scheme & Minimum  fairness & Target & Threshold & $\theta$ & Cost (mean $\pm$ se)\\  \midrule
POP & 75\% & HH & 0.2 &  56.23 & 168655 $\pm$ 14592 \\
POP & 90\% & HH & 0.4 & 74.98 & 176377 $\pm$ 14389 \\
POP & 99\% & HH LH & 0.8 & 56.23 & 293956 $\pm$ 20785  \\
NEB & 75\% & HH LH & 0.7 &  56.23 & 112870 $\pm$ 944 \\
NEB & 90\% & HH LH & 0.7 &  56.23 & \textbf{112870 $\pm$ 944} \\
NEB & 99\% & HH LH & 0.7 &  56.23 & \textbf{112870 $\pm$ 944} \\
NI-DEG & 75\% & HH & 0.005 &  17.78 & \textbf{66891 $\pm$ 2185} \\
NI-DEG & 90\% & HH LH & 0.007 &  23.71 & 143260 $\pm$ 2000 \\
NI-DEG & 99\% & HH & 0.017 &  31.62 & 512727 $\pm$ 1885 \\
NI-EIG & 75\% & HH & 0.003 &  31.62 & \textbf{66252 $\pm$ 2623} \\
NI-EIG & 90\% & HH & 0.003 &  74.98 & 190862 $\pm$ 3974\\
NI-EIG & 99\% & HH & 0.017 &  42.16 & 705906 $\pm$ 2352\\
\end{tabular}
\end{table*}
\end{center}

\subsection{Social diversity reduces interference complexity}

Social diversity introduces several challenges which must be overcome by an external decision-maker, but these bring with them opportunities to exploit the inherent mix of strategies that can be successful, according to different initial network conditions. If we consider a hierarchy of complexity based on the inherent costs of gathering information as explained above, we show that two targets have the potential to be optimal in a wide range of schemes and fairness requirements (see Table \ref{table:ug-ba-full}). Ensuring both offers and responses are fair is the strict, but also intuitive approach to investment. Nevertheless, several configurations in which rewarding fair responders (i.e., HH or LH) succeed as the most cost-effective avenues towards fairness. 

In the presence of diversity, we see that strictness, while generally effective, is not necessarily optimal. Heterogeneity allows for the coexistence of several strategies in a cluster, and relaxing the eligibility conditions for investment can allow an external decision-maker to reinforce positive behaviour, ultimately producing the optimal outcomes shown in Table \ref{table:ug-ba-full}. 
These results contrast starkly with previous observations in structured populations, for which stringent information gathering (i.e., targeting HH) leads to the lowest total costs and the highest levels of fairness, in almost all cases. Previously, for structured populations, relaxing targeting conditions was only desirable in the presence of high mutation rates. Diversity acts in a similar fashion to the noise associated with high mutation rates, which also allows the coexistence of several strategies. 
This is because heterogeneity allows for the coexistence of several strategies in a cluster, and relaxing the eligibility conditions for investment can allow an external decision-maker to reinforce positive behaviour, ultimately producing the optimal outcomes shown in Table \ref{table:ug-ba-full}. 

\begin{figure}
    \centering
    \includegraphics[width=\linewidth]{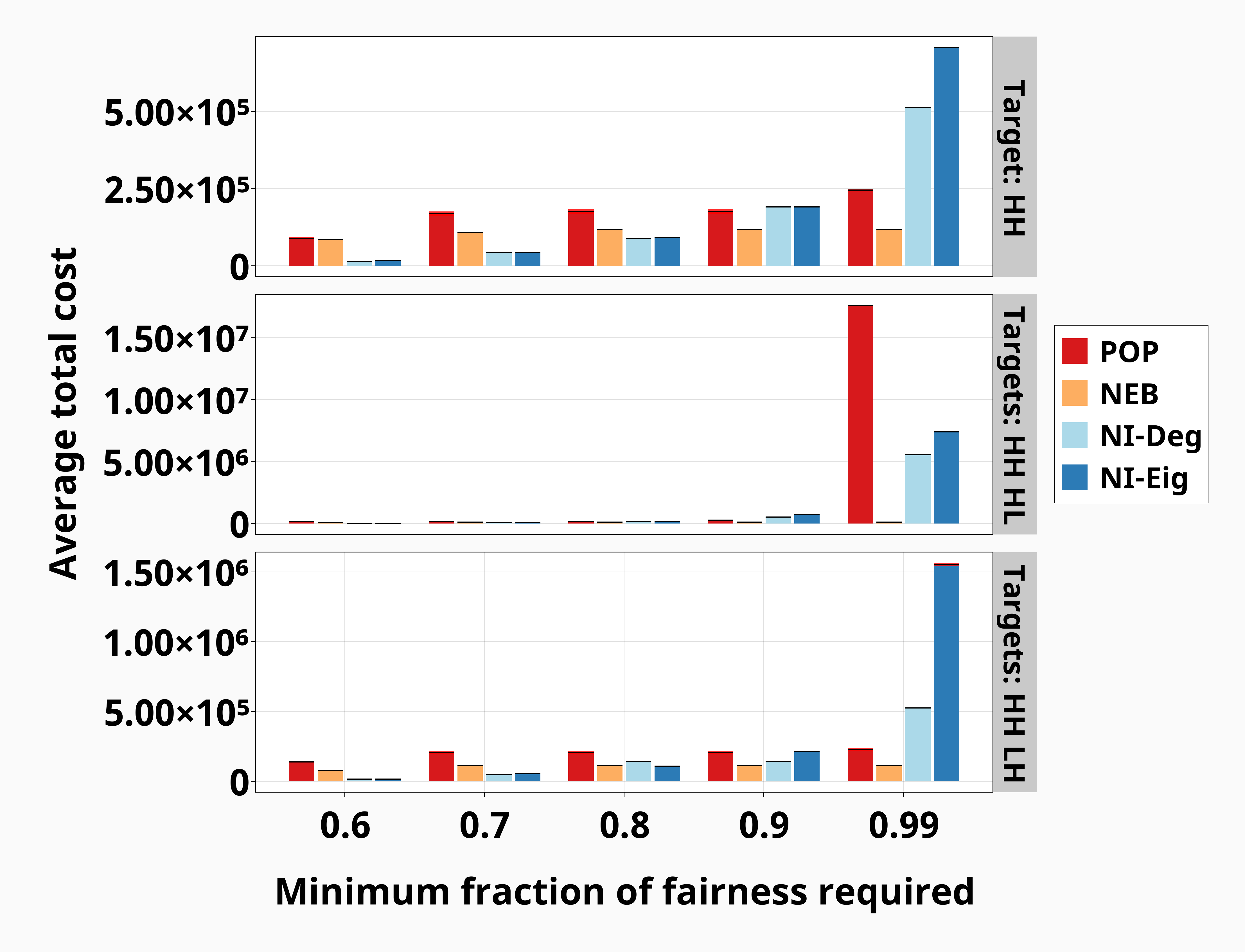}
    \caption[Mean total costs]{\textbf{Mean total costs for each possible target and scheme}. The figure illustrates the mean total costs for the most efficient combinations of threshold and investment amount $\theta$ for each possible target and scheme in BA networks. Error bars are shown in light red. Colours correspond to four different approaches: POP - population-based (red colour), NEB - neighbourhood-based (orange), and node influence-based by considering respectively degree centrality (NI-deg) (light blue) and eigenvector centrality (NI-eig) (blue).} 
    \label{fig:ug-barplots-ba}
\end{figure}

\subsection{Standards of fairness stipulate divergent approaches}

Fundamentally, institutional incentive schemes, in the context of asymmetric interactions and the Ultimatum Game, are diametrically different to cooperative dilemmas where heterogeneity naturally promotes pro-sociality  \citep{santos:2005:prl,santos2008social, cimpeanu2019exogenous}. As the UG allows for a larger room for improvement, due to relatively low baseline fairness levels (see again Section \ref{sec:ug} and Figure \ref{fig:ug-baseline-ba}), an investor can modify their goals and opt for modest improvements. They could, for instance, decide to invest only in large hubs, and ignore any potential outliers. Whether due to budget constraints, lack of information, or even uncertainty of network characteristics, they could adjust their margin for improvement and prevent unnecessary spending. 
Figure \ref{fig:ug-barplots-ba} shows the average total cost required for optimal investment schemes across a wide range of goals. We note the differences in the scales of the y-axes, which imply that on average, the strictest target (HH) is also the cheapest, followed by ensuring fair responses (HH and LH) and finally only ensuring fair proposals (HH and HL), which is significantly more expensive on average. This variance increases substantially with higher requirements for fairness. 

Figure \ref{fig:ug-barplots-ba} shows that targeting hubs (highly connected nodes) prevents over-spending as long as a small fraction of unfairness is accepted. Targeting hubs is the most cost-efficient strategy when the minimum fraction of fairness is low but this benefit decreases as the minimum fraction of fairness increase. In the case where less than 1\% of unfair individuals is accepted, the cost of targeting hubs is much higher than other interference strategies. This is because the hubs' spheres of influence do not extend far enough towards the leaves of the graph for them to make a marked improvement. Local observations, while comparatively expensive for most minimal fairness requirements, have the benefit of being able to extend their reach to lowly connected nodes.

Given a low enough minimum fairness requirement, targeting hubs (highly connected nodes) prevents over-spending. As these requirements increase, we observe that the hubs' spheres of influence do not extend far enough towards the leaves of the graph for them to make a marked improvement. Local observations, while comparatively expensive for most minimal fairness requirements, have the benefit of being able to extend their reach to lowly connected nodes. We further exemplify the effects of such an approach in Figure S4 in the SI, where we select one of the most cost-effective values of individual investment $\theta$ and show outcomes for all possible combinations of targets and centrality measures. 

Moreover, the differences between the two schemes (NI and NEB) are exacerbated for more demanding targets. Ensuring only fair responses implies the suitability of a swathe of nodes, thus being comparatively more effective at promoting fairness than the stricter target (HH), which restricts node candidacy. An external decision maker should be strict about which hubs to invest into, but can be more lenient in the selection of sparsely connected individuals. This is a promising result, as the cost of information gathering is assumed to scale with the number of subjects. Investors could potentially afford to spend the implied additional costs to scrutinise influential individuals, while allowing for a much broader classification if they opted for local observations, instead.

\subsection{Polarisation towards fairness}
\begin{figure*}
    \centering
    \includegraphics[width=1\linewidth]{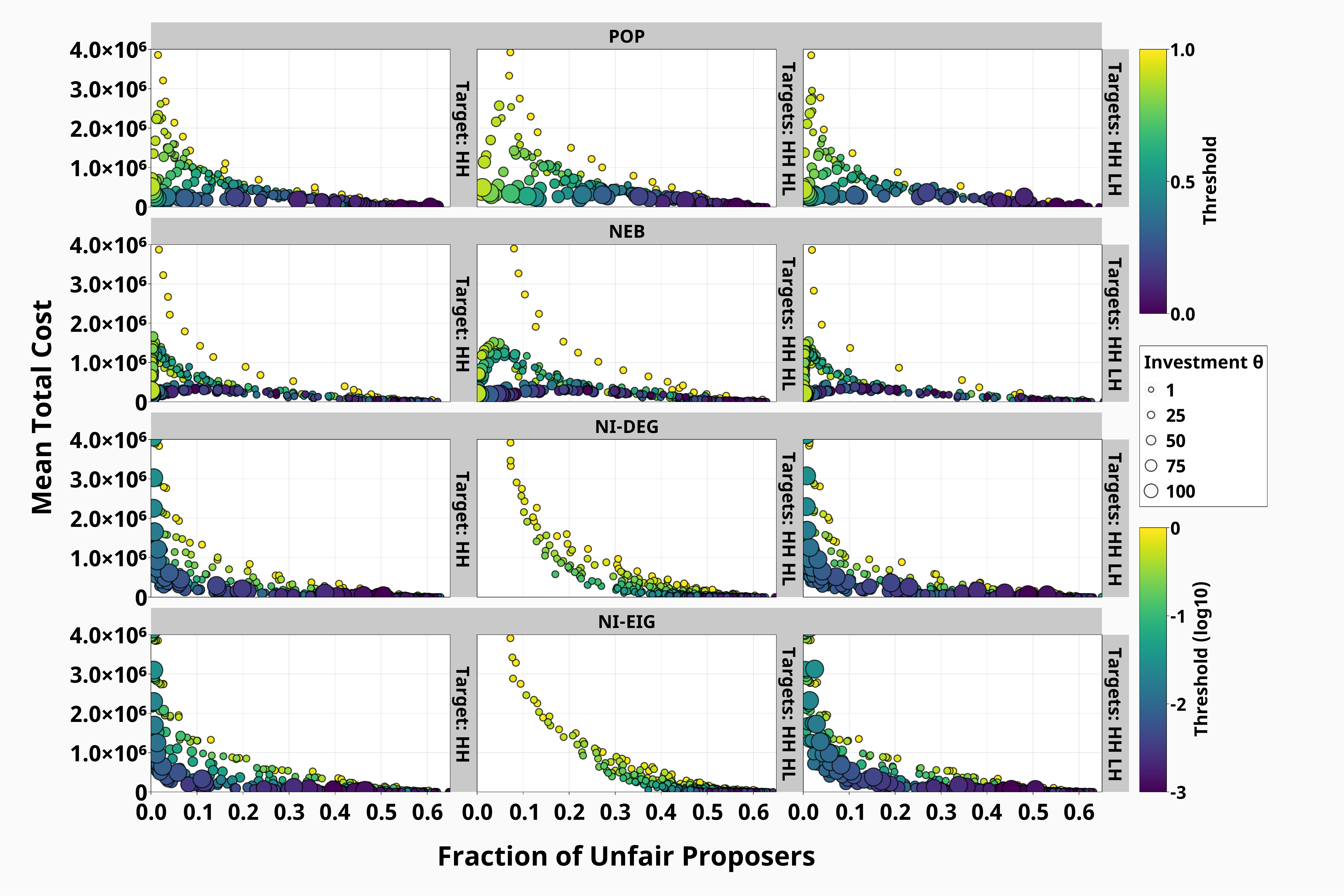}
    \caption[Pareto fronts for each scheme in BA networks]{\textbf{Pareto fronts for each scheme in BA networks}. We show the proportion of unfair proposers as a function of the average interference cost for each scheme and target combination. The markers' size is determined by the individual investment $\theta$ (grouped to the nearest value), whereas the colour is determined by the threshold (ranging from `blue' (lowest) to `yellow' (highest) value). Markers near the origin indicate the optimal solutions. Note that we only show the most cost-effective solutions, by limiting the maximum total cost.}
    \label{fig:ug-pareto-ba}
\end{figure*}

\begin{figure}
    \centering
    \includegraphics[width=0.9\linewidth]{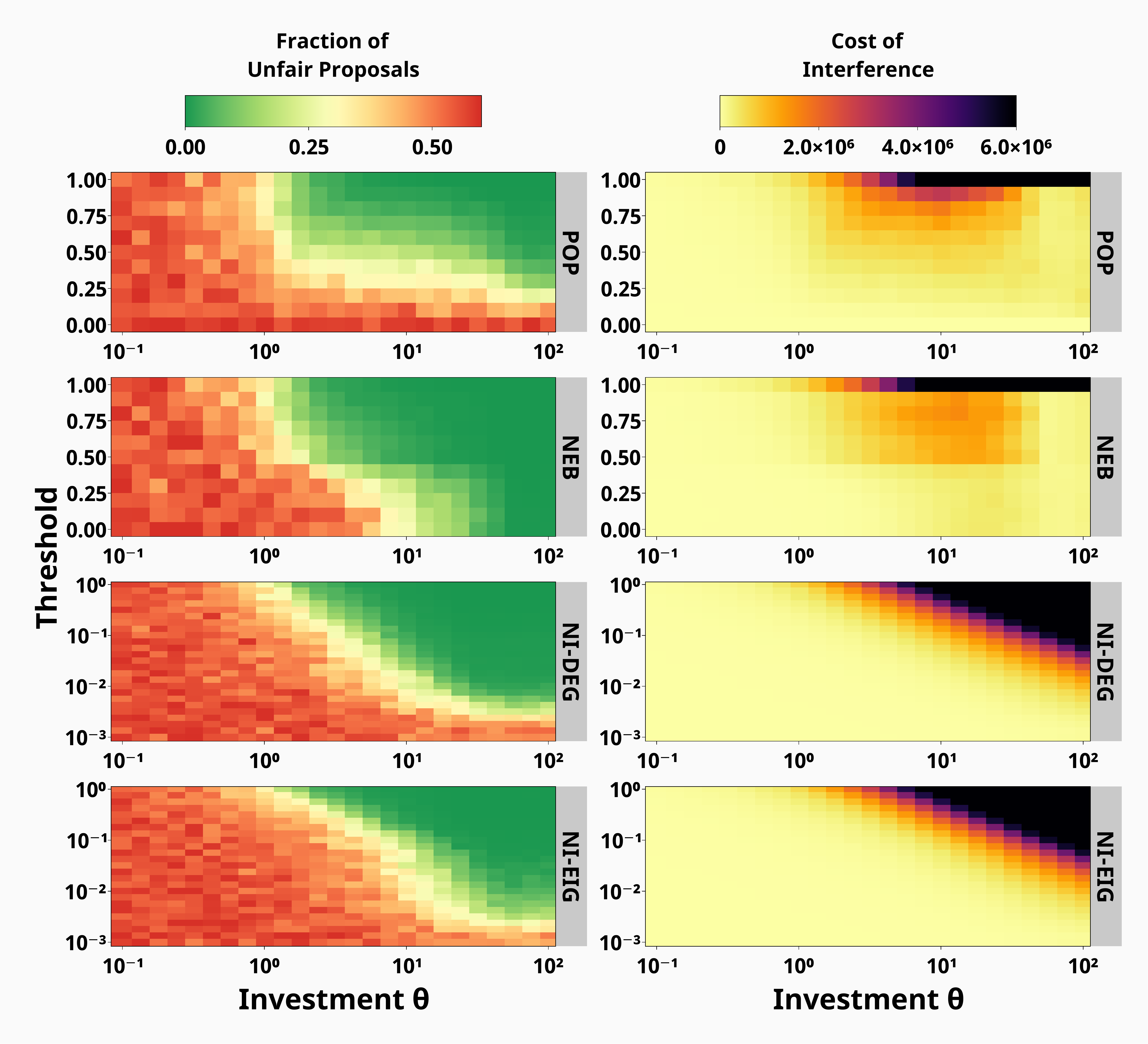}
    \caption[Targeting fair responders in BA networks]{\textbf{Targeting fair responders in BA networks}. The figure show the proportion of unfair proposers and total costs of investment for each scheme, while targeting fair responses (HH and LH). In the left column, colour corresponds to the proportion of unfair proposals, ranging from `green' (lowest) to `red' (highest) (see colour bar on the top of left column). In the right column, colour corresponds to the interference, ranging from `yellow' (lowest) to `black' (highest) (see colour bar on the top of right column).} 
    \label{fig:ug-heatmaps}
\end{figure}

Departing significantly from previously discussed results on homogeneous populations, we observe a tendency towards polarisation (see Figures \ref{fig:ug-pareto-ba} and \ref{fig:ug-heatmaps}). Across a wide spectrum of both interference schemes and various targets, we show that a very large $\theta$ can propel the system towards fairness, while also minimising cost. After having reached close to 100\% fairness, it is difficult for unfair strategists to invade the network, due to the heterogeneous dynamics at play, so no further investment is required.

By varying the threshold for investment, the amount of funding that is required for the system to shift towards fairness also changes. In a phenomenon akin to energy landscapes encountered in physical systems, the ``energy'' required to push fairness across the local maxima increases the further away it gets from the global minimum. Lowering the threshold can be beneficial, as overspending is avoided. On the other hand, the obstacle that arises is two-fold. As the goal is ultimately to reach a high level of fairness, a low threshold allows unfair individuals to thrive before fair individuals are eligible for investment. In other words, the system dips further away from the global minimum, and the investment amount required increases appropriately. 

Theoretically, this would allow for a number of viable investment approaches in such a system, but there are also several practical concerns of note. Institutions wishing to employ the practice of increasing their endowment amounts would be expected to have access to a considerable amount of initial funding, as opposed to spreading out the costs over multiple investment rounds. Moreover, there exists an intermediary region where overspending becomes problematic (see Figure \ref{fig:ug-heatmaps}). This issue occurs when the investment amount is large enough to be effective at inducing the change towards fairness, but not at the level where it does so rapidly. Given a relatively high threshold, many candidates become suitable for assistance, thus leading to excessive funds being deployed in locations of the network which are ineffective. 

\begin{figure} 
    \centering
    \includegraphics[width=1.\linewidth]{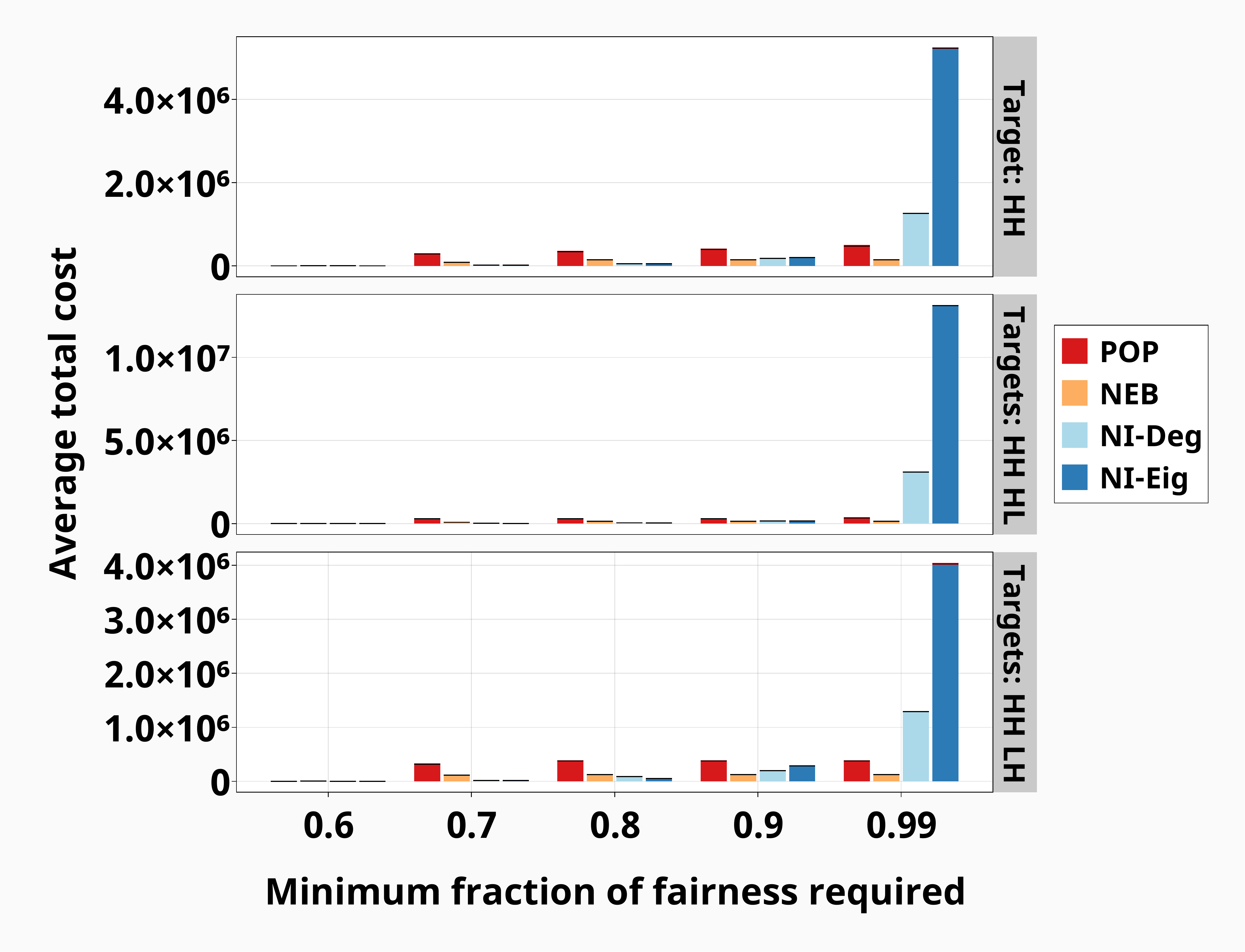}
    \caption[Most efficient schemes in DMS networks]{\textbf{Most efficient schemes in DMS networks}. Mean total costs (scaled by log10) for the most efficient combinations of threshold and investment amount $\theta$ for each possible target and scheme. Colours correspond to four different approaches: POP - population-based (`red' colour), NEB - neighbourhood-based (`orange' colour), and node influence-based by considering respectively degree centrality (NI-deg) (`light' blue colour) and eigenvector centrality (NI-eig) (`blue' colour). Error bars are shown in light red. If a certain scheme is missing, no investment was triggered for each desired standard of fairness.}
    \label{fig:ug-barplots-dms}
\end{figure}

\begin{center}
\begin{table}
\centering
 \caption[Most cost-efficient interference schemes to reach a minimum fairness of proposals in DMS networks]{\textbf{Most cost-efficient interference schemes to reach a minimum fairness of proposals in DMS networks.}} 
 \label{table:ug-dms-full}
\small
\begin{tabular}{c c c c c c}\toprule
Scheme & Minimum  fairness & Target & Threshold & $\theta$ & Cost (mean $\pm$ se)\\  \midrule
POP & 75\% & HH HL & 0.8 & 74.98 & 294102 $\pm$ 34137 \\
POP & 90\% & HH HL & 0.8 & 74.98 & 294102 $\pm$ 34137 \\
POP & 99\% & HH HL & 0.9 & 74.98 & 342927 $\pm$ 30915 \\
NEB & 75\% & HH LH & 0.3 & 74.98 & 123497 $\pm$ 7776 \\
NEB & 90\% & HH LH & 0.3 & 74.98 & 123497 $\pm$ 7776 \\
NEB & 99\% & HH LH & 0.3 & 74.98 & 123497 $\pm$ 7776 \\
NI-DEG & 75\% & HH LH & 0.004 &  10.00 & 25173 $\pm$ 1060 \\
NI-DEG & 90\% & HH HL & 0.004 &  42.16 & 158095 $\pm$ 1113 \\
NI-DEG & 99\% & HH & 0.031 &  42.16 & 1260950 $\pm$ 5993\\
NI-EIG & 75\% & HH HL & 0.001 &  17.78 & 14995 $\pm$ 376 \\
NI-EIG & 90\% & HH HL & 0.004 &  42.16 & 160555 $\pm$ 1343 \\
NI-EIG & 99\% & HH LH & 0.177 &  23.71 & 4026960 $\pm$ 32543 \\
\end{tabular}
\end{table}
\end{center}

\subsection{Clustering further reduces the burden on investors}

Real-world scale-free networks often have high transitivity (i.e., clustering), a feature missing in BA networks \citep{Su2016,barrat2005rate}, so it is crucial to measure the effects of network transitivity on the choice of investment policies. In the absence of investment, high transitivity positively influences fairness, thereby lowering the total amount of costs required to reach minimum standards of fairness (see Figure \ref{fig:ug-barplots-dms}). Moreover, eigenvector centrality (NI-Eig) can be employed to reduce overall spending for moderate fairness requirements, and hubs can be exploited to reduce spending for all but the most strict fairness regimes, unlike in lowly clustered networks. Comparing optimal investment schemes, we show that costs remain similar across all but the minimal desired standards of fairness (see Figure \ref{fig:ug-barplots-dms} and Table \ref{table:ug-dms-full}). Strict fairness regimes can be enforced using local information without overspending, similar to BA networks, yet we also show that population-based metrics are less risky in highly clustered networks. Thus, network transitivity acts as an equaliser between the different schemes and targets. 

Previously, we had seen that targeting proposers and responders (HH) and solely proposers (HH and HL) were both sensible approaches towards leveraging fairness. In the presence of clustering, targeting fair responders (HH and LH) becomes viable and in some cases, optimal (see Table \ref{table:ug-dms-full}). In fact, we see that diversity and transitivity act similarly to noise or behavioural exploration in the choice for investment, but on a much broader scale. Both of these factors open up novel mechanisms of engineering fairness while minimising the risk of choosing inappropriate candidates for endowments. With this reduction in complexity, we also see a very slight increase in the overall costs required to promote fairness. Paradoxically, a higher baseline of fairness also requires more endowments to reach fully positive outcomes. Heterogeneity acts as an equalizer in the truest sense, aiding in the quest towards a fair society, but also fostering a small minority of unfair individuals. We note that all the findings outlined above are consistent across both types of scale-free networks, and across interference schemes. For a detailed view of each scheme,  see Figures S5--S13 in  SI. 

\section{Discussion}
\label{sec:conclusions}

In this work, we have introduced social diversity in the Ultimatum game and studied the effects of heterogeneity on external interference, taking into account cost optimisation, limited information and standards of fairness. We found that diversity reduces the need for complex information gathering, and allows for less strictness in the eligibility criteria for receiving endowments. Exploiting different measures of centrality has been shown to enable a reduction in spending if standards of fairness are lowered. Our results indicate that the presence of diversity reduces the cost and complexity of promoting the evolution of fairness.

Existing models of institutional incentives aimed at promoting collective behaviours, such as cooperation and fairness \citep{Sigmund2001PNAS,han2022institutional,sasaki2012take,sun2021combination,sigmund2010social,gois2019reward}, usually ignore the problem of cost-efficiency.  These works often consider non-adaptive incentive mechanisms, studying how minimal incentive mechanisms can promote cooperation. The overarching goal of our research is to exploit available information that can be gathered to design  cost-efficient, adaptive incentive mechanisms. Our work also differs from EGT literature on optimal control in networked populations, where cost-efficiency is not considered \citep{ramazi2015analysis,riehl2016towards}. Instead, these works on controllability focus on identifying which individuals or nodes are the most important to control (i.e. where individuals can be assigned strategies as control inputs), for different population structures.

Previous works studying the optimisation of the cost of providing institutional incentives in an evolving population have mainly focused on  cooperative behaviours and symmetric games, namely the Prisoner's Dilemma (PD) \citep{han2018cost,han2018ijcai,cimpeanu2019exogenous,chen2014optimal} and Public Goods Games (PGG) \citep{chen2015first,wang2019exploring, DuongHanPROCsA2021}. 
Both homogeneous and heterogeneous networks have been analysed, showing that taking into account local network properties, such as detailed neighbourhood information and node degrees, can significantly reduce the cost required to ensure a certain level of cooperation. 
There are some recent efforts to extend these analyses to asymmetric games, namely the Ultimatum Game, but they are limited to the simpler setting of homogeneous networks, in the form of well-mixed and lattice graphs \citep{cimpeanu2021cost}. Thus, the present work advances this literature where the external decision maker needs to account for both the heterogeneous characteristics of agents and the hierarchical asymmetry in their interactions. 
As shown, cost-efficient interference mechanisms that incorporate this combined information can outperform those which only consider global population statistics and neighbourhood  properties \citep{cimpeanu2021cost}.

 In the context of institutional incentives modelling for promoting enhanced collective behaviours, an important issue is  how to set up and maintain the incentive budget. The problem of who  contributes to the budget is a social dilemma in itself, and how to escape it is a challenging research question. Facilitating solutions include pool incentives with second-order punishments \citep{sigmund2010social,perc2012sustainable}, democratic decisions \citep{hilbe2014democratic}, or mixed incentives (both positive and negative) \citep{chen2015first}, just to name a few. In this work, we do not address this issue, focusing instead on how to optimise the spending from a given  budget by exploiting network properties and information gathering. However, it would be interesting to study the co-evolutional institutional formation with different interference strategies and individual strategic behaviours.

Real-world networks, such as networks of collaboration and social networks, are often inherently \textit{heterogeneous}, whereby individuals differ in the number of connections, capacities, etc. \citep{barabasi2014linked}. The presence of individual and network structural heterogeneity has been demonstrated to be crucial in the evolution of various collective behaviours (e.g. cooperation, coordination, trust, or AI  safety) in networked games \citep{kumar2020evolution,santos2008social}. However, little attention has been given to how this heterogeneity affects external or institutional decision-making, e.g. to optimise the institutional cost of providing incentives to achieve high levels of such collective behaviours. The present paper's analysis aims to shed light on this issue, providing insights into how fairness can be efficiently engineered in real-world, dynamical settings (such as hybrid networked societies of human and autonomous systems) \citep{Andras2018TrustingSystems,dafoe2021cooperative}. 

Works on collective behaviour in social networks usually assume that changes are initiated from inside the system \citep{ranjbar2014evolution, FranksGA14}. Even when the role of influencers was explored \citep{FranksGA14}, external interference mechanisms and incomplete information are not taken into account. Mechanism design schemes have been employed to successfully resolve problems of incentivisation and taxation to enforce desired behaviours \citep{endriss2011incentive}, but these assume that the decision-maker has complete control over the agents within the systems. Our approach assumes that the decision-maker has little or no control over the agents, relying only on rewarding schemes, and nudging agents towards positive outcomes. Moreover, the literature on mechanism design typically does not focus on the costs of maintaining positive outcomes, whereas cost optimisation is one of our main objectives. 

In summary, we have shown that it is crucial to consider the roles' asymmetry to provide cost-efficient investment strategies, an important feature which was not possible to identify in previous works where symmetric games were studied \citep{chen2014optimal,han2018cost,han2018ijcai,chen2015first,wang2019exploring,cimpeanu2019exogenous, DuongHanPROCsA2021}. We have identified several key features that are required to minimise costs while ensuring positive outcomes. We found that diversity reduces the need for complex information gathering, and allows for less strictness in the eligibility criteria for receiving endowments. These results, regardless of the underlying interaction structure, stand out in sharp contrast with previous works on cooperation dilemmas, in which interference schemes require an exceptionally strict investment approach \citep{han2018ijcai,han2018cost}. 

\end{document}